# Metasurface-enabled quantum holograms with hybrid entanglement

Hong Liang, Wai Chun Wong, Tailin An, and Jensen Li[*]

*Department of Physics, The Hong Kong University of Science and Technology, Clear Water Bay, Kowloon, Hong Kong, 999077, P. R. China*

*\* jensenli@ust.hk*

**Abstract:** Metasurfaces, with their capability to control all possible dimensions of light, have become integral to quantum optical applications, including quantum state generation, operation, and tomography. In this work, we utilize a metasurface to generate polarization-hologram hybrid entanglement between a signal-idler photon pair to construct a quantum hologram. The properties of the quantum hologram can be revealed by collapsing the polarization degree of freedom of the idler photon, inducing interference between two holographic states of the signal photon, as a meaningful and selective erasure of the holographic content. In contrary, interference disappears when the idler photon is detected without observing polarization. This process can be further interpreted as a quantum holographic eraser, where the erasing action is visualized with erased contents in holograms. Our construction of polarization-hologram hybrid entangled state with metasurfaces will be useful for quantum communication with enhanced robustness, anti-counterfeiting applications through the additional quantum degrees of freedom, and as an emerging platform for exploring fundamental quantum concepts for entanglement and non-locality.

## 1. Introduction

Metasurface, composed of a thin layer of subwavelength structures, has a distinct capability to modulate light, particularly in hybridizing various degrees of freedom (DoFs) of light, such as wavelength, polarization, and orbital angular momentum (OAM) [1-6]. Recent advancements have significantly expanded the applications of metasurfaces towards the quantum regime, encompassing high-dimensional entangled photon state generation, quantum interference manipulation, quantum imaging, quantum state tomography, and quantum algorithms implementation [7-18]. A pivotal feature of quantum metasurfaces is their hybridization capability, facilitating the precise manipulation of quantum entanglement, which is a cornerstone of quantum information science [19-21]. Specifically, metasurfaces enable effective generation of quantum hybrid entanglement between different DoFs of photon pairs, such as entanglement between polarization and OAM or between polarization and path [22-24]. However, the exploration of quantum entanglement involving the spatial holographic field, one of the most complex DoFs of light, has been limited due to the challenging implementation of such entanglement using conventional bulky optical components. Metasurfaces become an ideal candidate to overcome this challenge, making spatial holographic entanglement accessible [25].

In this work, we utilize a metasurface and entangled photon pairs (signal and idler photons) to generate quantum holograms based on polarization-hologram hybrid entanglement. The metasurface in signal arm generates two distinct holographic states of the signal photons, which are entangled with two orthogonal polarization states of the idler photons, respectively. This hybrid entanglement enables us to remotely control the quantum holograms of the signal photon by adjusting the idler polarization. Specifically, we demonstrate a selective erasure of signal holographic contents when we insert a polarizer in the idler arm. This erasure results from the interference between the two tailored-made holographic fields generated by the metasurface, with the interference controlled remotely through the idler polarizer. Notably, this process can also be interpreted as a quantum eraser experiment [26-33]. The polarization of the idler photon acts as a path marker, revealing which-hologram-path information of the signal photon and thus its particle nature, leading to the absence of interference between two holograms. On the contrary, inserting a polarizer (eraser) in the idler arm removes this path marker, erasing the which-path information and revealing the wave nature of the signal photon, which restores interference between the two holographic paths. This erasing action is visualized as the selective erasure of holographic contents through remote control via different idler polarization selections. A notable benefit of utilizing holograms, recognized as high-dimensional spatial states, is the straightforward visualization of the quantum erasing action. Additionally, expanding quantum entanglement to incorporate spatial holograms holds promise for enhancing channel complexity and counterfeiting resilience in quantum communication [34-36]. Furthermore, such a metasurface-integrated approach to investigate quantum holograms marks a notable expansion for metasurface to explore fundamental quantum concepts – entanglement and quantum erasers. In addition to broadening the practical applications of metasurfaces, this work also enhances our insights into the core principles underlying quantum optics and information processing.

## 2. Quantum holograms based on metasurface-enabled hybrid entanglement

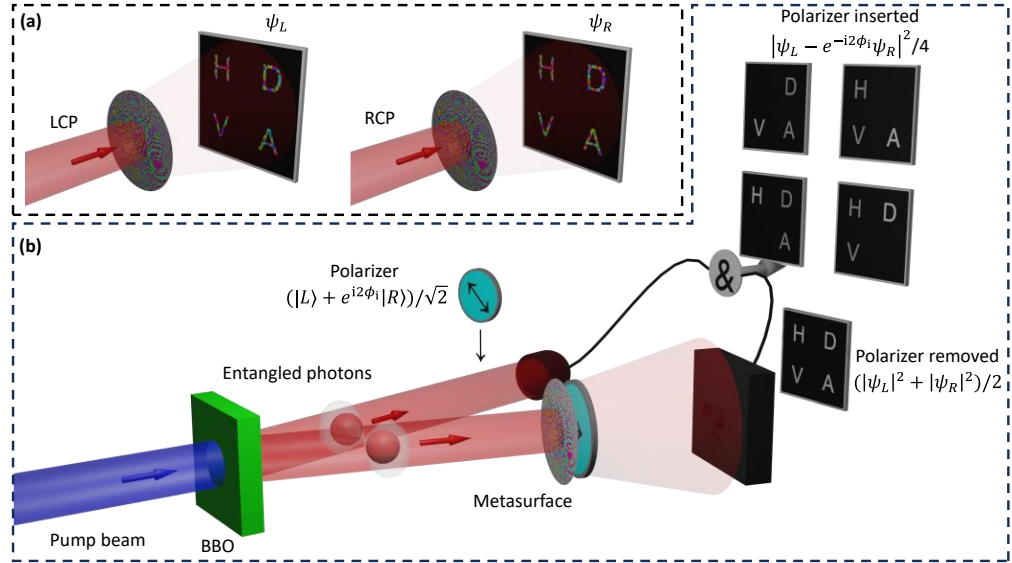

**Fig. 1. Quantum holograms based on metasurface-enabled hybrid entanglement.** (a) The two holographic states generated with LCP/RCP light incidence. The two holographic fields share the same amplitude distribution but have tailored relative phase relationship. (b) Schematic for polarization-hologram entanglement generation with an entangled photon pair and a geometric metasurface. A pair of entangled photons (idler and signal) are generated through a nonlinear crystal. The signal photon goes through a metasurface, generating the hybrid entangled state. Inserting and rotating the polarizer in the idler arm reveals different quantum holographic contents of the signal photon.

The proposed scheme to generate quantum holograms based on hybrid entanglement utilizes a polarization-entangled photon pair and a geometric metasurface, as shown in Fig. 1. The polarization state of the entangled photon pair (idler and signal photons) is $|\phi\rangle =$

$(|L\rangle_i|L\rangle_s - |R\rangle_i|R\rangle_s)/\sqrt{2}$, where the $|L\rangle$ ($|R\rangle$) stands for left (right) -handed circular polarization (LCP/RCP), and the subscript i (s) represents idler (signal) photon. The metasurface in the signal arm generates two polarization-dependent holograms at the observed plane with complex holographic field distribution $\psi_L(r)$ and $\psi_R(r)$ under the LCP and RCP light incidence respectively. These LCP and RCP holograms are assumed to be planar (2D) for simplicity and be scalar in specification, i.e, projected to a particular output polarization: horizontal polarization in this work. These two holograms, as seen in Fig. 1(a), are purposely designed with identical amplitudes, featuring four letters "HDVA", but differ in phase, which is specifically designed to enable the erasure of desired elements through the interference. Representing the basis holographic states with ket notation as $|\psi_L\rangle$ and $|\psi_R\rangle$, the metasurface can be effectively described as a spin-orbit coupling operation, denoted as $\hat{M}_s = |\psi_L\rangle_s\langle L|_s + |\psi_R\rangle_s\langle R|_s$, of which the detailed operation will be described later. Consequently, the resultant quantum state for the photon pair is

$$|\Phi\rangle = \hat{M}_s|\phi\rangle = \frac{1}{\sqrt{2}}(|L\rangle_i|\psi_L\rangle_s - |R\rangle_i|\psi_R\rangle_s), \quad (1)$$

indicating a construction of polarization-hologram hybrid entangled state, referred as the quantum hologram. To reveal the properties of the quantum hologram, we consider observing the signal arm with and without polarizer in the idler arm. A direct heralded measurement on signal photon's hologram without any polarization selection of idler photon (polarizer removed) reveals the incoherent mixture of the two holographic states as $\text{Tr}_i[|\Phi\rangle\langle\Phi|] = (|\psi_L\rangle\langle\psi_L| + |\psi_R\rangle\langle\psi_R|)/2$, giving an intensity distribution as $(|\psi_L|^2 + |\psi_R|^2)/2$, showing all four letters (Fig. 1(b)). On the other hand, inserting a polarizer (eraser) in the idler arm to select polarization collapses the signal photon to the corresponding superposition of holographic states. For example, as shown in Fig. 1(b), by projecting the idler photon to the linear polarization state of angle $\phi_i$: $|\xi\rangle_i = (|L\rangle_i + e^{i2\phi_i}|R\rangle_i)/\sqrt{2}$, the signal photon collapses to $_i\langle\xi|\Phi\rangle = (|\psi_L\rangle - e^{i2\phi_i}|\psi_R\rangle)/2$, which has an intensity field distribution of $|\psi_L - e^{i2\phi_i}\psi_R|/4$. With polarizers along different angle $\phi_i$, the resultant signal holographic field is targeted to illustrate different contents with the corresponding selected letter removed in the current example.

## 3. Metasurface design

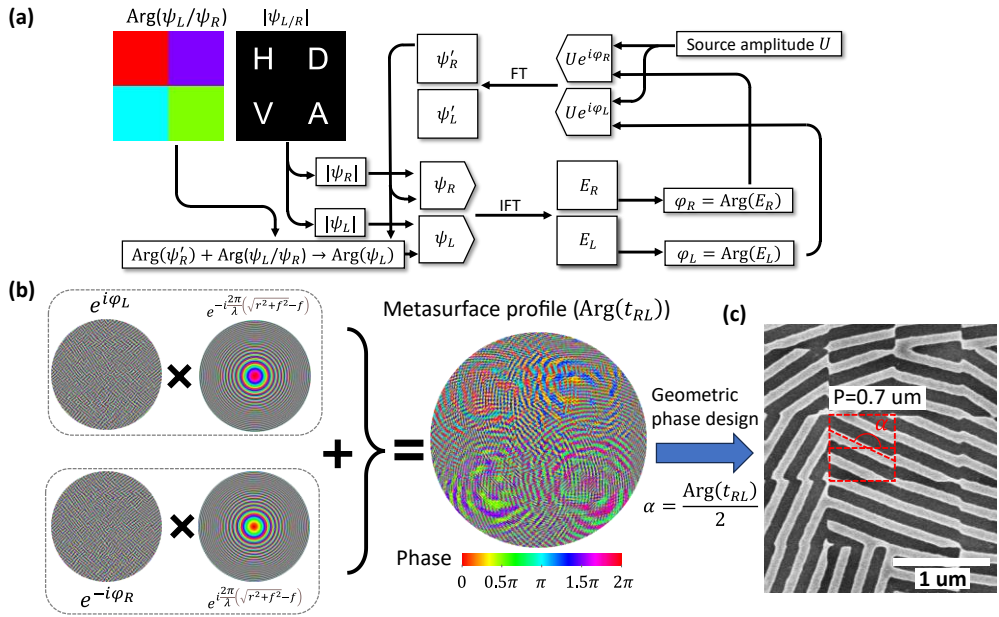

**Fig. 2. Metasurface design.** (a) The modified GS algorithm to design two phase profiles that generate two holograms with a designed phase relationship. The inputs are the target hologram amplitude $|\psi_{L/R}|$ and the phase relationship $\text{Arg}(\psi_L/\psi_r)$. The outputs are two phase mask profiles $\varphi_L$ and $\varphi_R$. (b) The metasurface profile for a geometric metasurface. The metasurface profile is obtained by combining $\varphi_L$ and $\varphi_R$ from the modified GS algorithm with converging lens and diverging lens profiles respectively. (c) The SEM image of the metasurface. The metasurface uses a geometric phase design. The period of the unit cell is 0.7 um and the rotation angle of the grating is half of the phase profile.

To construct the quantum hologram, we use a geometric metasurface to generate the LCP and RCP holograms with a common amplitude profile but different phase profiles at the image plane. The phase difference between the two holograms at various locations plays an important role in manipulating the interfered holographic result when a polarizer is inserted in the idler arm. Specifically, we adopt a modified Gerchberg–Saxton (GS) algorithm for designing the transmission phase profile of a metasurface. Without losing generality, an image with the letters "HDVA" is picked as our common hologram amplitude $|\psi_{L/R}|$, and the phase difference between the two holograms: $\text{Arg}(\psi_L/\psi_R)$ is picked to show four discrete values $0, 3\pi/2, \pi$, and $\pi/2$ for the regions of letters "H", "D", "V", and "A", respectively, as shown in the left hand side of Fig. 2(a). Then, the modified GS algorithm takes these as input and yields two phase masks $\varphi_L$ and $\varphi_R$ such that $\text{FT}(Ue^{i\varphi_{L/R}}) = \psi_{L/R}$, where FT represents the Fourier transform, and $U$ is the uniform source amplitude. The only modification of the GS algorithm is an additional constraint to ensure the phase difference $\arg(\psi_L/\psi_R)$ complies with specification. Up to this point, it seems that we need two metasurfaces to implement these two phase masks as transmission phase profiles. However, if we exploit the fact that these two holograms are for two orthogonal circular polarizations and we only collect the real image at a fixed distance $f$ (called focal length here) from the metasurface, one can actually combine the two phase masks to be implemented by the transmission phase profile, let's say $L$ to $R$ polarization, for a single geometric metasurface using

$$\text{Arg}(t_{RL}) = \text{Arg}\left(e^{i\varphi_L}e^{-i\frac{2\pi}{\lambda}\left(\sqrt{r^2+f^2}-f\right)} + e^{-i\varphi_R}e^{i\frac{2\pi}{\lambda}\left(\sqrt{r^2+f^2}-f\right)}\right), \qquad (2)$$

where the wavelength $\lambda$ is designed at 810 nm and the focal length $f$ is designed at 1000 μm. In this case, the LCP hologram is generated as a real image at the distance $f$ as $\psi_L$ for LCP incidence by noticing the converging phase profile $e^{-i\frac{2\pi}{\lambda}\left(\sqrt{r^2+f^2}-f\right)}$ added to it, while the RCP hologram is a virtual image by the diverging lens profile $e^{i\frac{2\pi}{\lambda}\left(\sqrt{r^2+f^2}-f\right)}$ and is not collected at the image plane. On the contrary, for RCP incidence, as the metasurface is assumed to compose of only geometric phase elements having $t_{LR} = t_{RL}^*$, the second term in Eq. (2) is turned into $e^{i\varphi_R}e^{-i\frac{2\pi}{\lambda}\left(\sqrt{r^2+f^2}-f\right)}$, meaning a real image $\psi_R$ at the image plane as the designed RCP hologram while the LCP hologram becomes diverging and is not collected. Such a scheme in designing the transmission phase of the metasurface from the two phase masks is summarized by Fig. 2(b). As a result, the metasurface operation can be effectively described as $\widehat{M}_{1,s} = |R, \psi_L\rangle_s\langle L|_s + |L, \psi_R\rangle_s\langle R|_s$, where the first slot of the ket represents output polarization state and the second slot represents spatial mode as a holographic state. Furthermore, to implement it using geometric phase elements, Fig. 2(c) shows the scanning electron microscope (SEM) image of our fabricated metasurface. Each unit cell with a period of 0.7 um consists of geometric-phase grating with a width of 105 nm and a rotation angle being half of the required transmission phase profile. We note that there is a residual part in the co-polarization channel, but it is not captured at the image plane since the converging lens phase profile only applies to the cross-polarization channel. For completeness, we have actually a final horizontal polarizer just after the metasurface, the operation now becomes $\widehat{M}_s = |H\rangle_s\langle H|_s\widehat{M}_{1,s} = 1/2(|H, \psi_L\rangle_s\langle L|_s + |H, \psi_R\rangle_s\langle R|_s)$, which can be simply written as $\widehat{M}_s = |\psi_L\rangle_s\langle L|_s + |\psi_R\rangle_s\langle R|_s$ with global factor and signal polarization state omitted for convenience, coupling the incident

light polarization and the generated hologram. Hereafter, we call this the operation of metasurface for simplicity.

## 4. Experimental demonstration

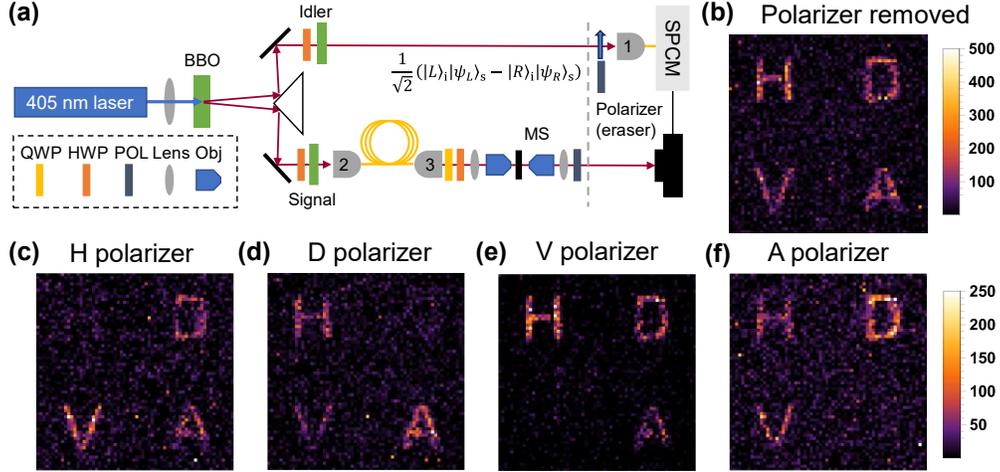

**Fig. 3. Experimental setup and results.** (a) The experimental setup. A 405 nm laser shining on a type-II β-barium borate (BBO) to generate polarization-entangled photon pairs, i.e., idler photon in the upper arm and signal photon in the lower arm. The signal photon interacts with the metasurface, and the generated hologram is imaged on the camera, which is correlated with the idler photon detection. (b) The signal photon's hologram shown without a polarizer (eraser) in the idler arm. (c-f) The holograms with different polarizers in the idler arm. The polarizer configured in horizontal (H), diagonal (D), vertical (V), and antidiagonal (A) positions erases the corresponding letter in the holographic results.

Figure 3(a) shows our experimental setup to generate and demonstrate the quantum hologram based on hybrid entanglement. Polarization-entangled photon pairs are first produced through spontaneous parametric down-conversion (SPDC) with a 405 nm pump laser (beam in blue color) on a type-II β-barium borate (BBO) crystal. The generated photon pairs are separated with a prism into the idler (upper) and signal (lower) arms. In each arm, a half-wave plate (HWP) with a 45° optical axis and a BBO with half the thickness of the main BBO are employed to compensate for the translational and longitudinal walk-off effects [37]. The generated state is conveniently expressed in terms of circular polarization as $|\phi\rangle = (|L\rangle_i|L\rangle_s - |R\rangle_i|R\rangle_s)/\sqrt{2}$. The signal photon is further directed through a 10-meter-long fiber for delay and then recollimated to free space. A quarter-wave plate (QWP) and an HWP before the metasurface are used to correct the minor polarization change due to the long-fiber transmission. With a lens and 10x objectives for imaging, the incident signal photon interacts with the metasurface and is imaged by a single-photon avalanche diode (SPAD) camera at a distance $f$ and with a horizontal polarizer placed before the camera. This setup results in the final quantum hologram, or the hybrid-entangled state for the photon pair at the dashed line in Fig. 3(a), being expressed as Eq. (1). Next, the property of the quantum hologram can be probed by collapsing the idler photon, which is used to herald the SPAD camera for imaging. The idler photon is detected by a single photon countering module (SPCM) with or without a polarizer in front of the detector. For the case without the polarizer in the idler arm, Fig. 3(b) shows the heralded hologram of the signal photons, where all four letters are visible. The intensity of the heralded hologram, $(|\psi_L|^2 + |\psi_R|^2)/2$, is obtained by simply adding the intensities for the two hologram channels without any interference. On the other hand, for the case of inserting a linear polarizer in the idler arm effectively introduces interference between the two holographic states.

Expressing the idler polarization state as $|\xi\rangle_i = (|L\rangle_i + e^{i2\phi_i}|R\rangle_i)/\sqrt{2}$, where $\phi_i$ represents the idler arm's linear polarizer angle relative to the horizontal axis, we can observe the heralded holographic state in the signal arm as

$$|\xi\rangle_s = {}_i\langle\xi|\Phi\rangle = \frac{1}{2}(|\psi_L\rangle_s - e^{-i2\phi_i}|\psi_R\rangle_s). \tag{3}$$

For example, in Fig. 3(c), the heralded hologram result erases the letter "H" when $\phi_i = 0$, i.e., a horizontal (H) linear polarizer is inserted. This occurs because the area corresponding to "H" shares the same phase in both holographic paths, as indicated by the red region in $\text{Arg}(\psi_L/\psi_R)$ in Fig. 2(a). Thus, the heralded holographic state $|\psi_L\rangle_s - |\psi_R\rangle_s$ is manifested with an erased letter "H" because of destructive interference. Here, for simplicity, the global factor $1/2$ is omitted in presentation. Similarly, with $\phi_i = \pi/4$ (diagonal (D) polarizer in the idler arm), the heralded hologram in the signal arm becomes $|\psi_L\rangle_s + i|\psi_R\rangle_s$. In this case, the letter "D" in the LCP hologram $|\psi_L\rangle$ exhibits a $3\pi/2$ phase shift over the same area in the RCP hologram $|\psi_R\rangle$, resulting in the erasure of the letter "D", as shown in Fig. 3(d). Figs. 3(e) and 3(f) display the corresponding results when $\phi_i$ is taken as $\pi/2$ and $3\pi/4$ with a vertical (V) and an antidiagonal (A) polarizer respectively, which erases the letters "V" and "A". Compared to the result obtained without the polarizer (Fig. 3(b)), the results with the polarizer exhibit selective erasure on corresponding letters. We note that we have purposely designed the heralded hologram shown up as selective erasure. We call the polarizer in the idler arm "eraser". The erased letters show a drop in intensity by -13.8 dB on average compared to the results with the eraser off, indicating effective erasure. Conversely, the remaining letters show an average Pearson correlation of 0.64 compared with the heralded hologram without erasure, with an average contrast of 7.5 dB (see details in Appendix C). This suggests that despite the selective erasure of certain letters, the remaining holographic content retains sufficient clarity and coherence, facilitating effective interpretation and analysis. Indeed, it is noteworthy that when a specific letter is erased due to total destructive interference, the opposite letter, which signifies the orthogonal polarization to the erased one, experiences enhancement with constructive interference as per the tailor-made design. Meanwhile, the other two letters are expected to display similar amplitudes, indicative of intermediate interference levels. For instance, in Fig. 3(c), the letter "H" is erased while the letter "V" appears brightest, with the letters "D" and "A" exhibiting lower and approximately the same intensity.

Interestingly, the above experiment in revealing the quantum hologram can also be interpreted as a quantum eraser now towards the holographic level, or called quantum holographic eraser. Here, the polarization-hologram hybrid entanglement enables the marking of which-hologram-path information of the signal photon by the polarization of idler polarization. Such a which-hologram marker's presence makes the two terms on the right-hand side of Eq. (1) orthogonal and thus the detection of the signal photon alone captures the particle nature of the signal photon with a density of state $(|\psi_L\rangle\langle\psi_L| + |\psi_R\rangle\langle\psi_R|)/2$. It is noted that no cross-term is in the result, indicating that the interference between $|\psi_L\rangle$ and $|\psi_R\rangle$ is destroyed because of the availability of which-hologram information. On the contrary, the measurement of the idler photon with a linear polarizer (eraser) inserted, erases the which-hologram information and thus restores the interference. This is interpreted as the wave nature of the signal photon with a superposition state of the two holographic states. Here, we have promoted the meaning for the quantum eraser to the hologram level and have purposely visualized the eraser action of the particle nature of photon as a selective erasure of the holographic content (the letters) via different polarization directions of the eraser.

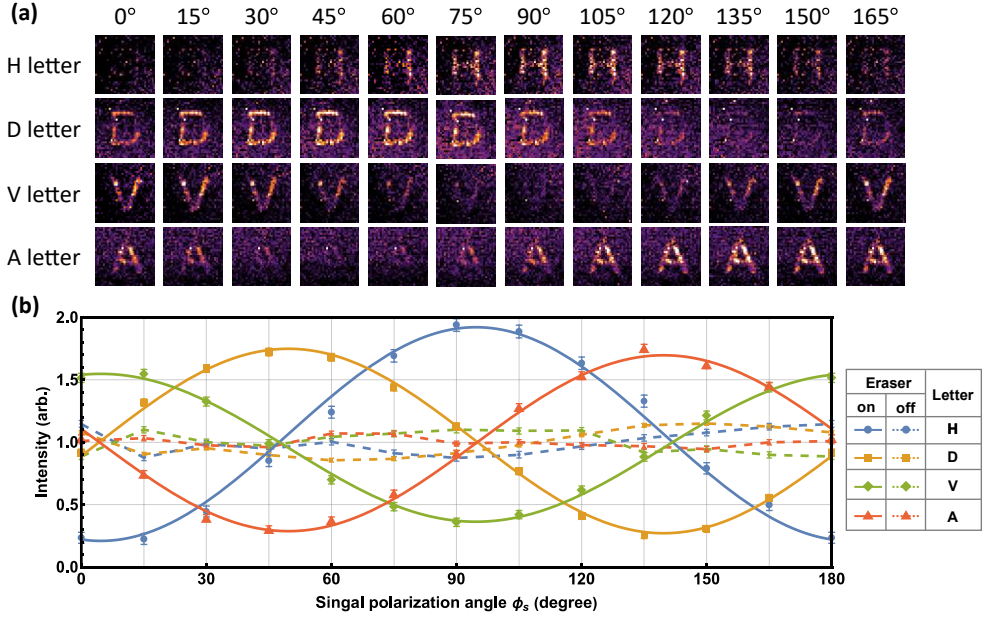

**Fig. 4. Quantum hologram interference with erasing effect.** (a) The interfered holographic image with different polarization projections in the signal arm when an H eraser is inserted in the idler arm. (b) Average intensity variation for individual letters with changing signal polarization angles when the eraser is inserted (on) and removed (off).

To further consolidate the existence of the interference between the two holographic paths, we now rotate the polarizer in the signal arm, which was previously fixed horizontally, with and without the polarizer (eraser) insertion in the idler arm. Fig. 4(a) shows that, with an H eraser in the idler arm, rotating the polarizer in the signal arm results in changing intensities for individual letters due to interference. On the contrary, without eraser insertion, the polarizer rotation in the signal arm would not affect the letters' intensities (as depicted in Fig. 5 in Appendix B), indicating the absence of an interference effect. Fig. 4(b) evaluates the average intensity of individual letters with signal polarizer at different angles when the eraser is on and off. When the eraser is on, the intensities appear as sinusoidal curves (solid lines), indicating the presence of interference between the two holograms. On the other hand, when the eraser is off, the intensities remain constant (dashed lines) because of the lack of interference.

In this case, we need to consider the polarizer in the signal arm. Taking the intermediate state before the signal polarizer $|\widetilde{\Phi}\rangle = \widehat{M}_{1,s}|\phi\rangle = 1/\sqrt{2}(|L\rangle_i|R,\psi_L\rangle_s - |R\rangle_i|L,\psi_R\rangle_s)$, we detect the idler photon to horizontal polarization and project the signal photon to different linear polarizations. The resultant holographic state can be expressed as $|\tilde{\xi}\rangle_s = 1/\sqrt{2}(_s\langle L| + e^{-i2\phi_s}{}_s\langle R|)_i\langle H|\widetilde{\Phi}\rangle = 1/2(e^{-i2\phi_s}|\psi_L\rangle_s - |\psi_R\rangle_s)$, where $\phi_s$ is the signal polarizer angle. Recalling that the two hologram fields have a spatial-dependent phase relation with each other, which can be expressed as $|\psi_L(r)\rangle = e^{i\theta(r)}|\psi_R(r)\rangle$, where $\theta(r) = \text{Arg}(\psi_L/\psi_R)$ is shown in Fig. 2(a), we can obtain the intensity distribution of the resultant heralded holographic state in the signal arm as $I(r) \propto |\psi(r)|^2|e^{-i2\phi_s}e^{i\theta(r)} - 1|^2/4 = |\psi(r)|^2\sin^2(\phi_s - \theta(r)/2)$, where $|\psi(r)| = |\psi_{L/R}(r)|$ as show in Fig. 2(a). We can further identify that there are four distinct regions in $\theta(r)$ showing different phases as $\theta_H = \theta(r \in "H") = 0$ (in the region of the "H" letter), and similarly $\theta_D = 3\pi/2$, $\theta_V = \pi$, $\theta_A = \pi/2$, where the subscript describes the corresponding regional letter. Consequently, the region with the letter "H" displays the intensity as $I_H \propto \sin^2(\phi_s - \theta_H/2) = \sin^2(\phi_s)$, as confirmed by the first row in Fig. 4(a) and the blue solid line with a visibility of 80% in Fig. 4(b), increasing from $\phi_s = 0$ to $\pi/2$ and decreasing afterward. Similarly, the intensity for the letter "D" is obtained as $I_D \propto \sin^2(\phi_s - \theta_D/2) = \sin^2(\phi_s - 3\pi/4)$, as shown by the second row in Fig. 4(a) and the yellow solid line with a visibility of 73% in Fig. 4(b), reaching its maximum and minimum at $\phi_s = \pi/4$ and $3\pi/4$. Furthermore, for the letters "V" and "A", the intensities are $I_V \propto \sin^2(\phi_s - \pi/2)$ and $I_A \propto \sin^2(\phi_s - \pi/4)$, shown as the third and fourth rows in Fig. 4(a), and evaluated with the green and red solid lines in Fig. 4(b), with visibilities of 62% and 71% respectively. It is noted that the fitting results show a global shift of around 4.6° for the four visibility curves due to the experimental errors from the angle of polarizers. On the other hand, when the eraser is off, rotating the signal polarizer results in the same image as $(|\psi_L|^2 + |\psi_R|^2)/2$, indicating unchanged intensity for all four letters. This is evaluated with four dashed lines in Fig. 4(b), with raw images shown in Fig. 5 in Appendix B. It is worth noting that we have used a GS algorithm modified to design and capture tailor-made interference between two holograms. Alternatively, it is also possible to resort to machine learning techniques to further improve the quality of the holograms with fewer speckles [38].

## 5. Conclusion

We have demonstrated the significant potential of metasurfaces in quantum optics by generating and manipulating quantum holograms through polarization-hologram hybrid entanglement. By employing a metasurface, we created a hybrid entanglement between the holographic state of a signal photon and the polarization state of an idler photon. The properties of the resultant quantum hologram can be revealed via the idler polarization selection, leading to selective erasure of holographic contents through interference in the signal arm. Our experimental success on constructing the quantum hologram highlights the practical and theoretical impacts of integrating high-dimensional holographic states into quantum entanglement. Photons with tailor-designed structured fields have the potential not only to enable more complex and secure communication channels but also to exhibit robustness against perturbative transmission and generate resilient quantum entanglement [ 39 - 41 ]. These intriguing features are highly beneficial for applications in quantum communication, teleportation, and information processing. Moreover, our experimental results also vividly illustrate the quantum eraser concept, demonstrating the interplay between the wave and particle nature of photons. We envision that the wave-particle duality can serve as an additional layer of protection in anti-counterfeiting applications in that a casual observation only reveals the particle nature, concealing the selective information underlying the wave nature, which is discovered only by using a specific idler polarizer. Our interpretation of the quantum holographic eraser also further opens new avenues for studying fundamental quantum phenomena, such as entanglement, non-locality, and the role of information in quantum optics. In a broader context, the integration of metasurfaces in quantum optics paves the way for compact, integrated quantum devices, crucial for the practical implementation of quantum technologies. Our study significantly advances the understanding and application of quantum metasurface holograms, indicating the transformative potential of metasurfaces in quantum optical applications.

## APPENDIX A: Materials and Methods

### Sample fabrication

The dielectric metasurfaces were fabricated on a quartz substrate following the processes of deposition, patterning, and etching. First, a 300 nm polycrystalline silicon (Poly-Si) film was deposited by low-pressure chemical vapor deposition (LPCVD). The inverse nanostructures were patterned using an E-Beam lithography system (JEOL JBX-6300FS) within 200 nm electron beam resist (CSAR 6200.09), followed by baking at 150°C for 1 minute. Finally, the desired structures were etched using SF6 and C4F8 in an inductively coupled plasma reactive-ion etching process (ICP-RIE).

### Quantum optical measurement
In the experiment, we use a 2-mm-thick type-II BBO and a 200mW 405 nm laser (CrystaLaser DL-405-400) to generate the polarization-entangled photon pairs in a state of $(|L\rangle_i|L\rangle_s - |R\rangle_i|R\rangle_s)/\sqrt{2}$. The half-opening angle of the generated photon pairs is designed to be 3°. The photons are split into the signal arm and idler arm using a prism. The photons in the idler arm are detected with polarization selection using a single photon counting module (SPCM) (Excelitas-SPCM-800-14-FC), and the detection signals are sent to herald signal photons' arrival on the SPAD camera. Meanwhile, the signal photons are sent through a 10-meter-long single-mode fiber to the imaging setup. A lens with a focal length of 75 mm and a 10x objective are used to focus the signal beam onto the metasurface. The hologram generated from the metasurface is imaged by the SPAD camera (SPAD512S) using a 10x objective and a lens with a focal length of 125 mm. The heralded image of the signal photons in Figs. 3(b) to 3(f) are all retrieved using 600 frames with external triggers from SPCM and with background white noise subtracted. The background is measured using the same triggering setting with blocked signal photons. Each frame spans 100 ms with a maximum of 255 photon counts in each pixel. Each trigger from the SPCM would turn on the camera for a detection window of 18 ns. With multiple triggers within one frame, the detection events in each triggered window would accumulate into one frame.

## APPENDIX B: Experimental results without erasure
Here Fig. 5 shows the quantum holographic results with a rotating signal polarizer without the erasure effect, which show no variation in the intensity level and are evaluated as the dashed curves in Fig. 4(b).

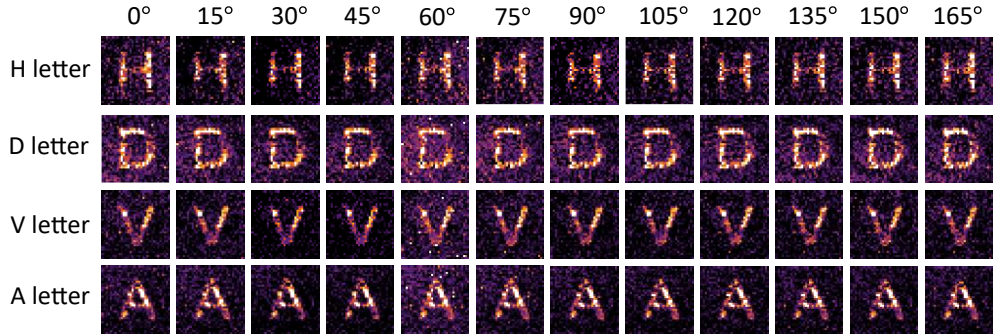

Fig. 5 Quantum hologram without the erasure effect.

## APPENDIX C: Calculation of intensity drop, contrast, and Pearson correlation

To evaluate the erasure effect on the holographic results, we calculate the intensity drop, Pearson correlation, and contrast on the images. The intensity drop is defined as

$$\text{Intensity drop}_{i,j} = 10 \log_{10}\left(I_{i,j}^{(e)}/I_i^{(w)}\right),$$

where $I_{i,j}^{(e)} = I_{i,j}^{(e,l)} - I_{i,j}^{(e,b)}$ is the intensity of individual letter $i \in \{H, D, V, A\}$ of the hologram result $j$ (a total of 4 results as shown in Figs. 3(c-f)) with erasure effect and is calculated as the difference between the average photon count of the letter $I_{i,j}^{(e,l)}$ and the background $I_{i,j}^{(e,b)}$ regions indicated as white and black regions in Fig. 6. Similarly, $I_i^{(w)} = I_i^{(w,l)} - I_i^{(w,b)}$ is the intensity of the letter $i$ of the hologram result without erasure. Then on average over the four holograms with erasure, the erased letters show a drop in intensity by -13.8 dB.

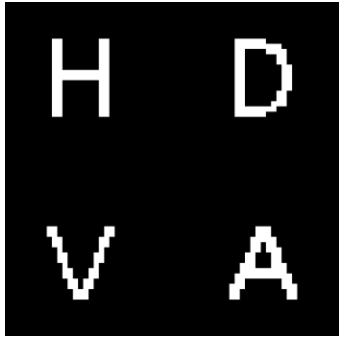

Fig. 6 The target holographic image as a mask to select letter and background regions.

For the remaining letters, we define the contrast for individual letters against the background as

$$\text{Contrast}_{i,j} = 10 \log_{10}\left((I_{i,j}^{(e,l)} - I_{i,j}^{(e,b)})/I_{i,j}^{(e,b)}\right),$$

and obtain an average contrast for the remaining letters as 7.5 dB. We further calculate the Pearson correlation coefficient for the remaining individual letter $i$ between the result $j$ with erasure and the result without erasure as

$$r_{i,j} = \frac{\sum_{x,y}(I_{i,j}^{(e)}(x,y) - \bar{I}_{i,j}^{(e)})(I_i^{(w)}(x,y) - \bar{I}_i^{(w)})}{\sqrt{\sum_{x,y}\left(I_{i,j}^{(e)}(x,y) - \bar{I}_{i,j}^{(e)}\right)^2}\sqrt{\sum_{x,y}\left(I_i^{(w)}(x,y) - \bar{I}_i^{(w)}\right)^2}},$$

where for example, $I_{i,j}^{(e)}(x,y)$ is the intensity distribution for letter $i$, including its regional background, in the hologram result $j$ with erasure. In this way, the remaining letters in all 4 results with erasure show an average Pearson correlation coefficient of 0.64, demonstrating a high correlation. The above analysis suggests that despite the selective erasure of certain letters, the remaining holographic content retains sufficient clarity and coherence, facilitating effective interpretation.

**Funding.** The work is supported by the Hong Kong RGC (16304020, 16306521, C6013-18G, AoEP-502/20) and by Croucher Foundation (CAS20SC01).

**Author contributions.** J.L. initiated the project. J.L. and H.L. conceived the idea. H.L., W.C.W. designed, and T.A. fabricated the metasurface. H.L. conducted the experiment and analyzed the data. H.L. wrote the manuscript, and J.L. and W.C.W. revised the manuscript. J.L. supervised the project.
**Disclosures.** The authors declare no competing financial interests.
**Data and materials availability.** All data needed to evaluate the conclusions in the paper are presented in the paper. Additional data related to this paper may be requested from the authors.